\documentclass[footinbib,twocolumn,showpacs,amsmath,amstex,amssymb,mathfonts,superscriptaddress,prl]{revtex4}
\usepackage{graphicx}
\usepackage{color}
\usepackage{bm}
\usepackage{amsmath}
\usepackage{amssymb}
\usepackage{amsthm}
\usepackage{amsfonts}

\def\qv{\vec{q}}

\newcommand{\norm}[1]{\left\lVert#1\right\rVert}
\usepackage{bbm}

\begin{document}

\title{Generalized Pseudopotentials for the Anisotropic Fractional Quantum Hall Effect}
\author{Bo Yang} 
\affiliation{Complex Systems Group, Institute of High Performance Computing, A*STAR, Singapore, 138632.}
\author{Zi-Xiang Hu}
\affiliation{Department of Physics, Chongqing University, Chongqing, P.R. China, 401331.}
\author{Ching Hua Lee}
\affiliation{Material Science and Engineering, Institute of High Performance Computing, A*STAR, Singapore, 138632.}
\author{Z. Papi\'c}
\affiliation{School of Physics and Astronomy, University of Leeds, Leeds LS2 9JT, UK}
\pacs{73.43.Lp, 71.10.Pm}

\date{\today}
\begin{abstract}
We generalize the notion of Haldane pseudopotentials to anisotropic fractional quantum Hall (FQH) systems which are physically realized, e.g., in tilted magnetic field experiments or anisotropic band structures. This formalism allows us to expand any 
translation-invariant interaction over a complete basis, and directly reveals the intrinsic metric of 
incompressible FQH fluids. We show that purely anisotropic pseudopotentials give rise to new types of bound 
states for small particle clusters in the infinite plane, and can be used as a diagnostic of FQH nematic order. We also demonstrate that generalized pseudopotentials quantify 
the anisotropic contribution to the effective interaction potential, which can be particularly large in models of fractional Chern insulators.
\end{abstract}

\maketitle 

The fractional quantum Hall (FQH) system is host to a wide variety of topological phases of matter \cite{Tsui-PhysRevLett.48.1559}.    
This complexity belies the deceivingly simple microscopic Hamiltonian containing only the effective Coulomb interaction projected to a single Landau level (LL) \cite{Prange}. The understanding of different topological states was greatly facilitated by the concept of \emph{pseudopotentials} (PPs) introduced by Haldane \cite{Haldane83, prangegirvin}. This formalism allows one to expand any rotation-invariant interaction over the complete basis of the PPs, which are projection operators onto two-particle states with a given value of relative angular momentum. Furthermore, a combination of a small number of PPs naturally defines parent Hamiltonians for some FQH model states, such as the  Laughlin states \cite{Laughlin-PhysRevLett.50.1395,TrugmanKivelson}. The method has also been generalized to many-body PPs \cite{SimonRezayiCooper, CHLeePapicThomale}, which form the parent Hamiltonians of the non-Abelian FQH states \cite{Moore1991362,ReadRezayiParafermion}. In many cases, the ground state of these model Hamiltonians is believed to be adiabatically connected to the actual ground state of the experimental system. Thus, the relatively simple (and to some degree analytically tractable) model wavefunctions and Hamiltonians give much insight into the nature of the experimentally realized FQH states.

Recently, interest in the FQH effect has been renewed due to emerging connections between topological order, geometry and broken symmetry. An early precursor of these ideas was the realization that rotational invariance is not necessary for the FQH effect \cite{prangegirvin}. This lead to the conclusion that FQH states possess new ``geometrical" degrees of freedom \cite{HaldaneGeometry}, uncovering a more complete description of their low-energy properties \cite{CanLaskinWiegmann, BradlynRead, Gromov}. The notion of geometry has also inspired the construction of a more general class of Laughlin states with non-Euclidean metric \cite{QiuPhysRevB.85.115308}, which were shown to be physically relevant in situations where the band mass or dielectric tensor is anisotropic \cite{BoYangPhysRevB.85.165318,  XinWanPhysRevB.86.035122, Apalkov2014128}, or in the tilted magnetic field \cite{PapicTilt}. On the other hand, an intriguing co-existence of topological order with broken symmetry \cite{Xia2011, Mulligan}, leading to the ``nematic" FQH effect, has also been proposed \cite{MaciejkoPhysRevB.88.125137, YouPhysRevX.4.041050}. The nematic order is expected to arise due to spontaneous symmetry breaking, as suggested by recent numerical evidence \cite{Regnault2016} and experiment \cite{Samkharadze2016}.

A missing ingredient for the complete microscopic description of the mentioned geometry and broken symmetry in the FQH effect is the formulation of PPs for systems without rotational invariance. A ``rotationally invariant" system is one where the effective interaction $V(q_x,q_y)$, which encompasses the bare Coulomb potential and the LL form factor \cite{Prange}, is a function of a single metric $g$, i.e., $V(q_x,q_y)=V(|q|=\sqrt{g^{ab}q_aq_b})$. Note that the metric $g$ does not necessarily have to be Euclidean in the lab frame, because its physics remain unchanged upon a trivial coordinate transformation back to the Euclidean metric. By contrast, if the Coulomb potential and LL form factor are characterized by different metrics, rotational symmetry will be `truly' broken. Note that a finite system can break rotational invariance by its boundary conditions, but recent work \cite{CHLeePapicThomale} has shown that conventional PPs can nevertheless be defined in such cases. In contrast, here we focus on describing a thermodynamically large system without rotational invariance. Such systems are highly relevant for experiment, yet they cannot be described by the conventional PP formalism. 

In this paper, we extend Haldane's approach \cite{Haldane83} and construct a complete basis of the generalized PPs for \emph{any} two-body effective interaction.
An immediate benefit of the generalized PPs is that anisotropic FQH systems, such as tilted magnetic field experiments, can be studied in a quantitative and universal language. Moreover, the generalized PPs lead to several physical consequences for FQH systems. First, we show that ground states of purely anisotropic PPs feature new types of bound states with ``molecular" structures. In combination with isotropic PPs, such molecular clusters might give rise to novel types of order in partially filled LLs. Second, we show the anisotropic PPs are a natural probe of the simplest FQH nematic states  \cite{MaciejkoPhysRevB.88.125137, YouPhysRevX.4.041050}, and motivate the construction of a more general class of orders that extend beyond the ``quadrupolar" instability of neutral excitations of the Laughlin state. Third, we explicitly evaluate the anisotropic PPs in a model fractional Chern insulator (FCI), and show they are comparable in magnitude to the isotropic PPs. This suggests that anisotropic PPs are not only physically relevant in FCIs, but can be useful for determining the optimal FCI models for realizing various topological states. Finally, we highlight some broader connections between generalized PPs and the emergent geometry of FQH systems.

{\it Generalized pseudopotentials.--}
Any two-body FQH Hamiltonian, projected to a single LL, is given by:
\begin{eqnarray}\label{h}
\mathcal H=\int \frac{d^2 \qv \ell_B^2}{\left(2\pi\right)^2} V_{\qv}\bar\rho_{\qv}\bar\rho_{-\qv},
\end{eqnarray}
where $\bar\rho_{\qv}=\sum_i e^{iq_aR_i^a}$, $a=\left(x,y\right)$, is the guiding-center density operator with momentum $\qv=\left(q_x,q_y\right)$ (we assume the Einstein summation convention). The model in Eq.(\ref{h}) is believed to explain most of the observed FQH states depending on the subtle energetics dictated by the electron filling fraction $\nu$ and the effective interaction potential $V_{\qv}$. The complexity of $\cal H$ in Eq.~(\ref{h}) is hidden in the fact that the guiding center coordinates do not commute: 
$[R_i^a, R_j^b]=-i\epsilon^{ab}\ell_B^2\delta_{i,j}$, where $\epsilon^{ab}$ is the anti-symmetric tensor and $\ell_B$ is the magnetic length (we set $\ell_B=1$). 

Interaction $V_{\qv}$ can be written as $V_{\qv}=V^{\text{bare}}_{\qv}|F_N\left(\qv\right)|^2$, where $V^{\text{bare}}_{\qv}$ is the 
bare Coulomb potential and $F_N\left(\qv\right)$ the form factor of the electrons in the $N^{\text{th}}$ LL. In the simplest case of a sample with infinitesimal thickness in the perpendicular direction, $F_N\left(\qv\right)=F_N\left(|q|_l\right)$ and $V^{\text{bare}}_{\qv}=V^{\text{bare}}_{|q|_c}$, with $|q|_{l(c)}^2 =q_aq_bg_{l(c)}^{ab}$.  The LL metric $g_l$ parametrizes the effective mass tensor, and the Coulomb metric $g_c$ describes the shape of the equipotential lines around the electrons in a dielectric \cite{HaldaneGeometry}. In the following, we first assume isotropic metrics $g_l=g_c=g=\mathbbm{1}_{2\times 2}$, and return to the more general case  in Eq.~(\ref{eq:metric}) at the end.

For two particles, the center of mass guiding center coordinates commute with the Hamiltonian of Eq.(\ref{h}). Thus, the relevant Hilbert space is spanned by two-particle eigenstates $|m\rangle$ of relative angular momentum $m$ \cite{prangegirvin}. Assuming $\Delta m \equiv (m'-m) > 0$, we obtain the following expression for the matrix element:
\begin{eqnarray}\label{element1}
\langle m|\bar\rho_{\vec q}\bar\rho_{-\vec q}|m'\rangle=\sqrt{\frac{m!}{m'!}}\left(i\sqrt{2}\textbf{q}^*\right)^{\Delta m}e^{-\frac{1}{2}|q|^2}L_m^{\Delta m}\left(|q|^2\right).
\end{eqnarray}
Here $m,m'$ are necessarily even for bosons and odd for fermions, $\textbf{q} \equiv \frac{1}{\sqrt{2}}\left(q_x+iq_y\right)$, and $L_m^{n}\left(x\right)$ is the generalized Laguerre polynomial. From Eq.(\ref{h}) it is clear that if $V_{\vec q}$ is rotationally invariant with $g_l=g_c=g=\mathbbm{1}_{2\times 2}$,  then $\langle m|\mathcal H|m'\rangle\sim\sum_nc_n\delta_{m,n}\delta_{m',n}$. It is thus useful to decompose the interaction into the standard Haldane PPs $V_m\left(|q|\right)$ as
$V_{|q|}=\sum_{m=0}^{\infty}c_mV_m\left(|q|\right)$, where $V_m\left(|q|\right)=e^{-\frac{1}{2}|q|^2}L_m\left(|q|^2\right)$. Many of the well-known FQH states are described by model Hamiltonians formed by linear combinations of select $V_m$.

\begin{figure}[htb]
\includegraphics[width=\linewidth]{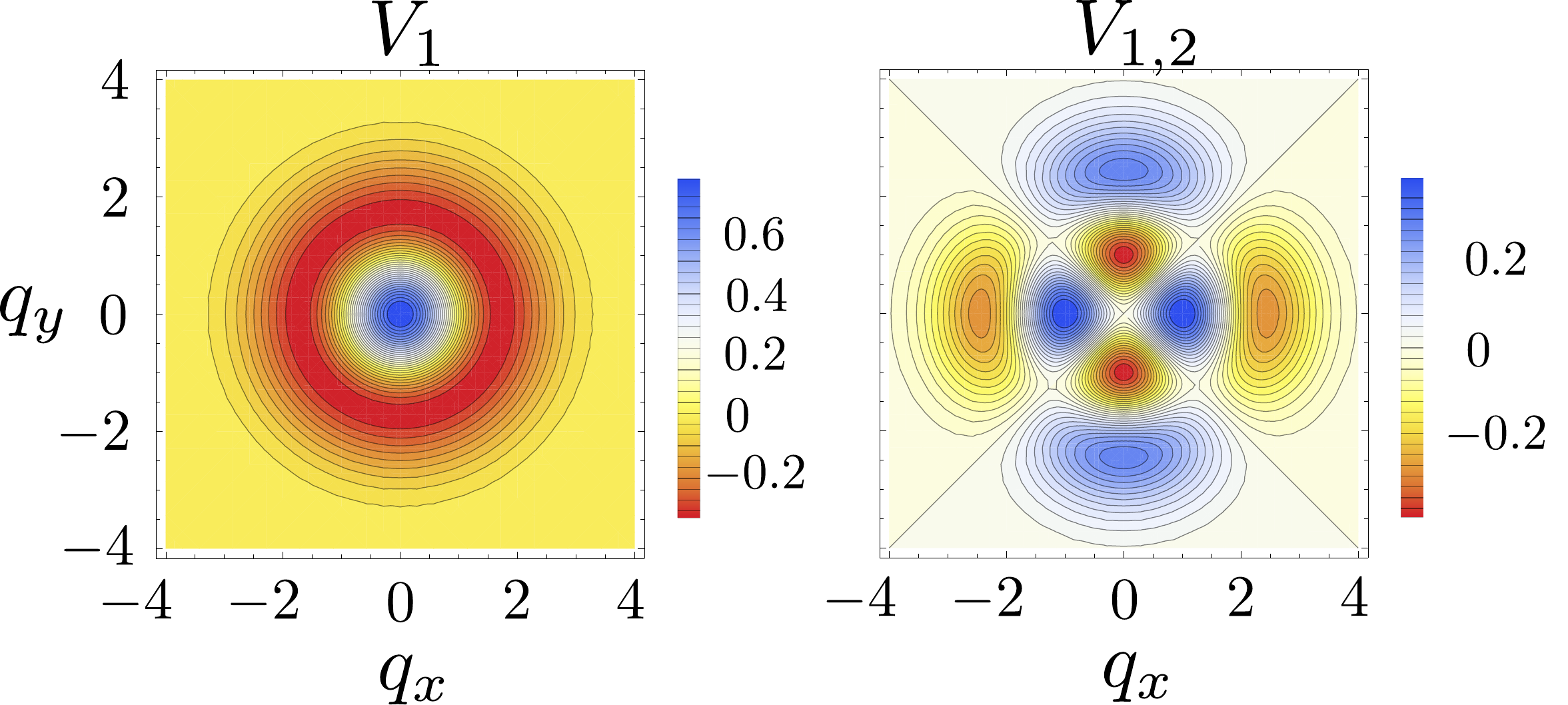}
\caption{ (Color online) Contours of the leading-order fermionic $V_1$ (isotropic) and $V_{1,2}^+$ anisotropic pseudopotential.} 
\label{fig:vmn}
\end{figure}

We next focus on arbitrary anisotropic $V_{\vec q}$. In this case, the Haldane PPs no longer form a complete basis. We define the \emph{generalized PPs} as follows: 
\begin{eqnarray}
V_{m,n}^{+}\left(\vec q\right)&=& \lambda_n \mathcal{N}_{mn} \left(L_m^n\left(|q|^2\right)e^{-\frac{1}{2}|q|^2}\textbf{q}^n+c.c\right),\label{g1}\\
V_{m,n}^{-}\left(\vec q\right)&=& -i \mathcal{N}_{mn} \left(L_m^n\left(|q|^2\right)e^{-\frac{1}{2}|q|^2}\textbf{q}^n-c.c\right),\label{g2}
\end{eqnarray} 
where the normalization factors are $\mathcal{N}_{mn}\equiv \sqrt{2^{n-1}m!/(\pi\left(m+n\right)!)}$, and $\lambda_n=1/\sqrt{2}$ for $n=0$ or $\lambda_n=1$ for $n\neq 0$. Physically, for $n\neq 0$, $V_{m,n}^{+}$ is equivalent to $V_{m,n}^{-}$ up to a rotation of $\frac{\pi}{2n}$ in  momentum space. Different generalized PPs are mutually orthogonal:
$\int d^2qV_{m,n}^{\sigma}\left(\vec q\right)V_{m',n'}^{\sigma'}\left(\vec q\right)=\delta_{m,m'}\delta_{n,n'}\delta_{\sigma,\sigma'}$,
and any effective interaction can be expanded as
\begin{eqnarray}
V_{\vec q}&=&\sum_{\substack{m,n=0\\ \sigma=\pm} }^\infty  c^{\sigma}_{m,n}V_{m,n}^{\sigma}\left(\vec q\right), 
c^{\sigma}_{m,n} = \int d^2q V_{\vec q}V_{m,n}^{\sigma}\left(\vec q\right).
\label{decompose2}
\end{eqnarray}
It is thus clear that we recover the Haldane PPs as special cases of $V_{m,n}^{+}$ with $n=0$ (and $V_{m,0}^{-}=0$), so we can write $V_{m,0}^{+}=V_m,c_{m,0}^{+}=c_{m},c_{m,0}^{-}=0$. For fermions, the shortest-range generalized PP is $V_{1,n}$; the quadrupolar contour profile of $V_{1,2}$ is clearly visible when compared against the isotropic $V_1$ Haldane PP (Fig.\ref{fig:vmn}). 

\begin{figure}
\includegraphics[width=0.9\linewidth]{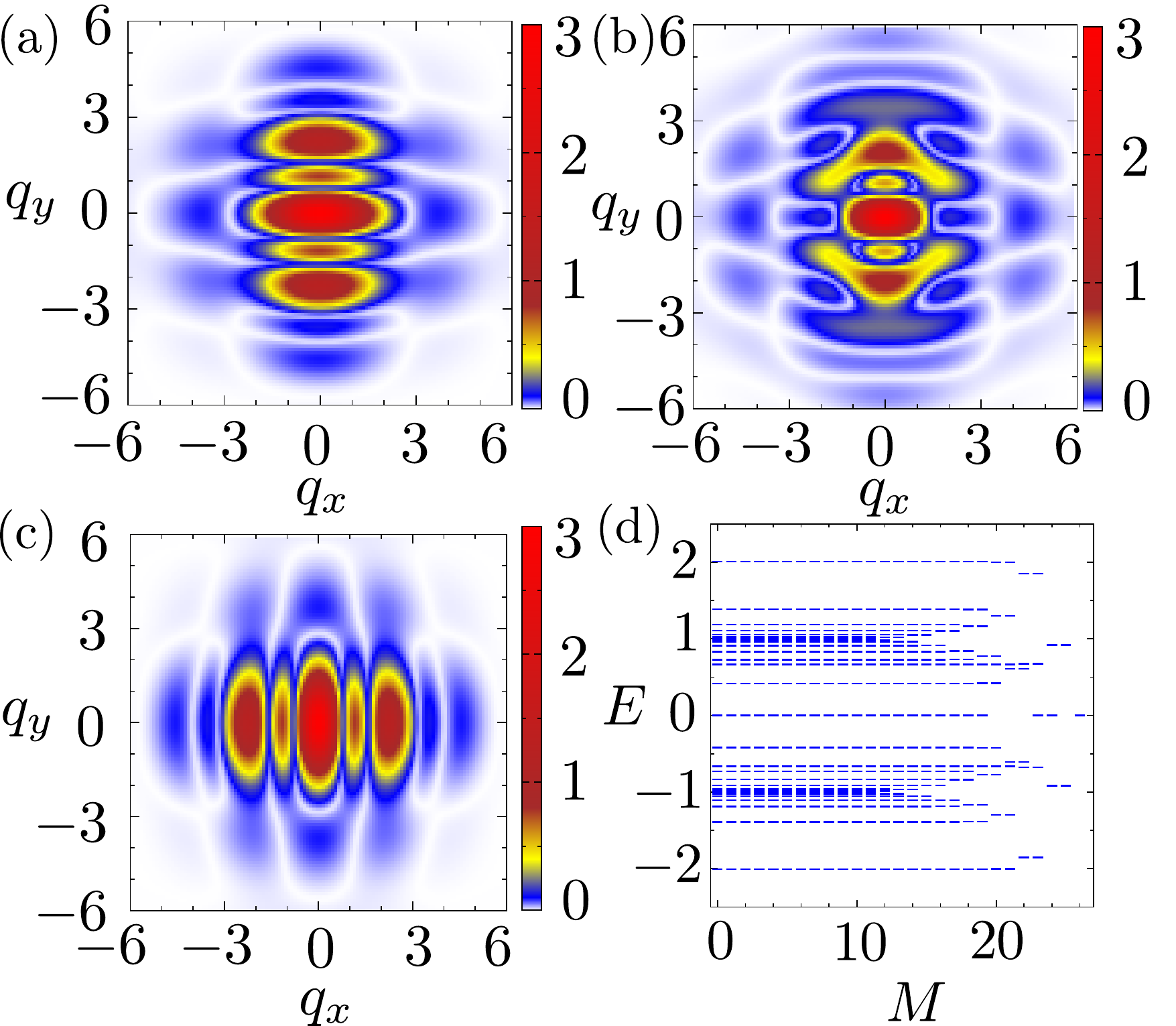}
\caption{(Color online) The guiding-center density $\langle\Psi|\bar \rho_{\vec{q}}|\Psi\rangle$ (a-c) and energy spectrum (d) of a 3-electron molecule on the disk with 30 magnetic orbitals. The density is evaluated in the ground state (a), first-excited state (b), and the highest-energy ``anti-bonding" state (c).The states are obtained by exact diagonalization of $V_{\vec q}=V_{1,2}^{+}$ in $M=0$ sector.} 
\label{fig3}
\end{figure}

{\it Intra-LL ``molecules"--} To gain some intuition about the generalized PPs, it is instructive to study the spectrum of the simplest anisotropic PP -- $V_{1,2}^{+}$ -- for small numbers of electrons on an infinite disk. 
From Eq.(\ref{element1}) it is clear that the ground state is a bound state with negative energy. The bound states can be understood as ``molecular levels" of these particles, and the guiding-center density profile of such ``molecules" in the center of mass (guiding-center) angular momentum $M=0$ sector are plotted in Fig.(\ref{fig3}). 

Note that for anisotropic Hamiltonians the effect of the disk boundary cannot be removed for any finite system size. This means that $M$, in contrast to rotationally-symmetric systems, is no longer a good quantum number for any eigenstate in the presence of the boundary. For bound states in a large disk, however, the particle density decays exponentially towards the boundary, so that the diagonalization within sub-Hilbert spaces of small $M$ are excellent approximations. The guiding-center density plots in Fig.(\ref{fig3}) thus represent very well the ``molecular structures" of the few particle states in an infinite plane at $M=0$. For three electrons, the ground state reflects an interesting bonding structure between the particles. There is also a localised ``anti-bonding" state at the highest energy, and the spectrum as a function of $M$ is symmetric about the zero energy. These properties are qualitatively the same for any small cluster, and will be studied in detail elsewhere. 

One unique feature of particles within a single LL (whose kinetic energy is quenched) is that the bound states exist even for purely repulsive interaction. For example, take
two particles with interaction
$ 
V_{\vec q}=V_{1}+\lambda V_{1,2}^{+},
$ The ground state of 
this system is a bound state regardless of how small $\lambda$ is. One can easily construct an effective interaction $V_{\vec q}$ consisting of $V_{0}, V_{1}, V_{2}$ and $V_{1,2}^{\pm}$, such that $V_{\vec q}>0$ for all $\vec q$. Eq.(\ref{h}) has bound states even when the interaction is repulsive everywhere, and the properties of these bound states are qualitatively the same as those in Fig.(\ref{fig3}), with the ``molecule" size increasing as $\lambda$ decreases. This unique feature is a direct consequence of the quenched kinetic energy and particle statistics, and does not apply to free particles in two-dimensional space, where no bound states exist for repulsive interactions.

{\sl Physical applications.--} We now provide two applications of the generalized PPs. First, we show that generalized PPs are a natural probe of the FQH nematic states \cite{MaciejkoPhysRevB.88.125137, YouPhysRevX.4.041050}. 
Second, we demonstrate that anisotropic PPs emerge naturally in FCIs \cite{Jackson2015}, where they can be comparable in magnitude to the isotropic PPs.  

The FQH nematic \cite{BalentsLiqudCrystal, Musaelian, Mulligan2010, Mulligan2011, MaciejkoPhysRevB.88.125137, YouPhysRevX.4.041050} is a phase with topological order and a charge gap, but at the same time spontaneously breaking rotational symmetry like a liquid crystal \cite{deGennes}. Recent interest in such phases is fuelled by experiment \cite{Xia2011,Samkharadze2016}, and effective field theories \cite{MaciejkoPhysRevB.88.125137,YouPhysRevX.4.041050} which describe nematic states as the long-wavelength instability of the quadrupolar spin-2 excitation of the Laughlin state \cite{HaldaneGeometry}. On the other hand, identifying nematic phases in microscopic models has proven more subtle. Recent work \cite{Regnault2016} suggested that a nematic phase might be realized in a model containing $V_1$, $V_3$ and $V_5$ Haldane PPs. However, the nematic phase was diagnosed by its response to changes in the sample geometry, which modifies all the interaction PPs at the same time, thus not conclusively ascertaining the quadrupolar character of the phase.    
\begin{figure}[htb]
\includegraphics[width=\linewidth]{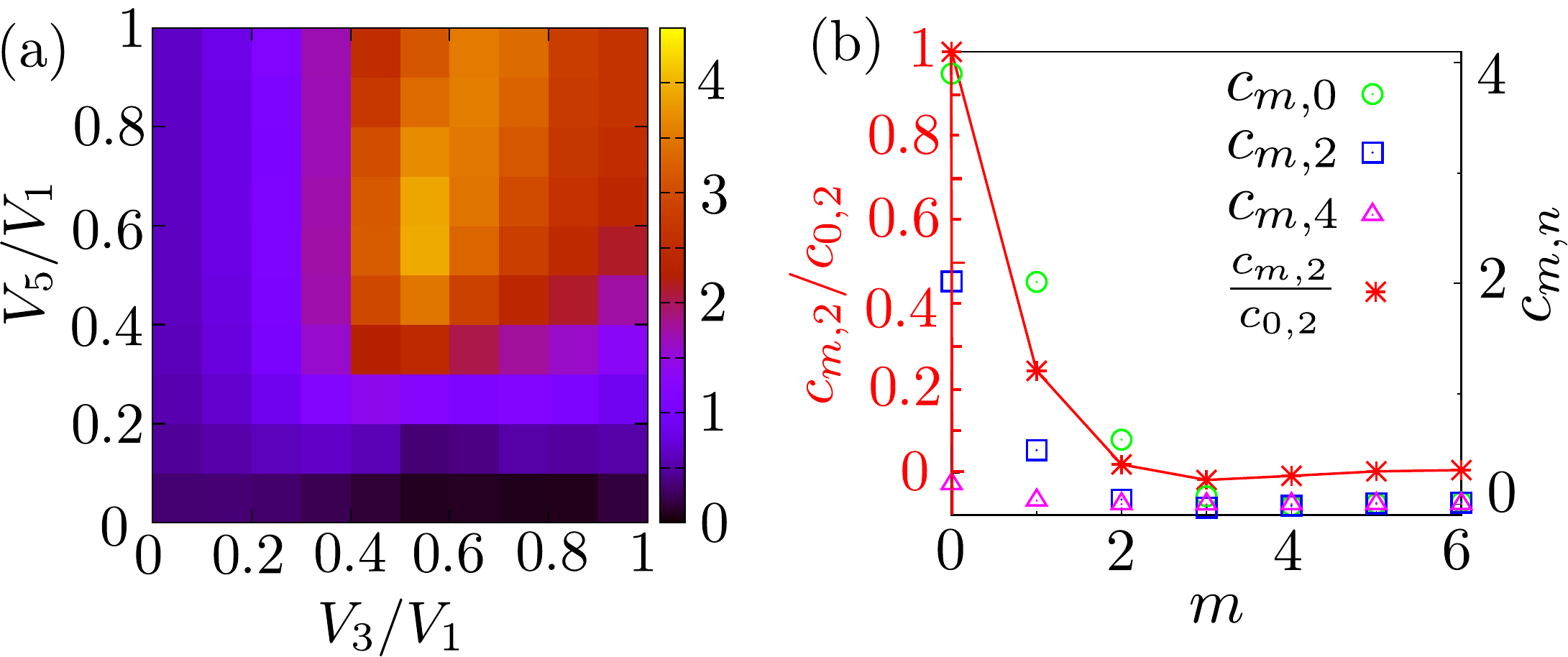}
\caption{(Color online) (a) Nematic susceptibility $\chi$ in a model with isotropic $V_1=1$, $V_3$, and $V_5$ Haldane PPs. The results are averaged over 50 lowest states of a system of 8 electrons on a square torus. The putative nematic phase \cite{Regnault2016} occurs around $V_3\approx V_5 \approx 0.6$.
(b) Anisotropic PPs in an FCI model. Left: the coefficients of anisotropic PPs for a quadrupole interaction $V_q=\cos q_x-\cos q_y$, whose isotropic PPs vanish. The first two PPs dominate. Right: PP magnitudes $c_{m,0}$,$c_{m,2}$,$c_{m,4}$ for $V_q=\cos q_x$. $V_{m,2}$ is of comparable magnitude as the isotropic PP $V_{m}$, at least for $m=0$ and $1$.
}
\label{fig4_v3}
\end{figure}  

The generalized PPs can be used as a diagnostic of the nematic phase by measuring the response to small perturbations of the isotropic potential by $V_{m,n}$, while preserving the geometry of the sample. In Fig. \ref{fig4_v3}(a) we plot the nematic response under adding or subtracting a small amount of $\delta V_{1,2}\sim 10^{-3}$. The nematic susceptibility plotted in Fig. \ref{fig4_v3}(a) is defined as
$
\chi 
\equiv 
| \mathcal{N}(V_m;\delta V_{1,2}) - \mathcal{N}(V_m;-\delta V_{1,2}) |
/(2\delta V_{1,2})
$, where the nematic order $\mathcal{N}$ is given by \cite{Regnault2016} $\mathcal{N} \equiv \frac{1}{N_\Phi}\sum_{\vec{q}}S_{\vec{q}} (\cos q_x - \cos q_y)$, in terms of the guiding-center structure factor, $S_{\vec{q}}$ \cite{Prange}. The data is for 8 electrons at filling $\nu=1/3$ ($N_\Phi=24$ flux quanta) on a square torus. Because of the gapless nature of the nematic phase, the response was averaged over 50 lowest eigenstates. It is clear that the nematic response strongly peaks around $V_3\approx V_5 \approx 0.6$; the location of the peak is in agreement with Ref. \cite{Regnault2016}, while at the same time the finite-size effect is reduced.

Another advantage of the method presented here is that instead of $V_{1,2}$, we can controllably vary other PPs such as $V_{1,4}$, $V_{3,2}$, etc. Interestingly, the nematic susceptibility under $\delta V_{1,4}$ is strongly suppressed  on the scale of Fig. \ref{fig4_v3}(a). The response under $\delta V_{3,2}$ is qualitatively similar to $\delta V_{1,2}$, but  smaller in magnitude. These results suggest the quadrupolar nature of the nematic phase, resulting from the instability of the long-wavelength part of the Laughlin magnetoroton mode \cite{GMP85}. This is also confirmed by the variational calculations using the magneroton wavefunctions \cite{BoYang2012}, which show the neutral excitation going soft at small momenta in the nematic region of Fig. \ref{fig4_v3}(a). It would be interesting to construct generalized nematic states whose primary instabilities occur in octupolar and higher order channels. 

As a second application, we show that large anisotropic PPs naturally arise in FCIs
 \cite{sheng2011fractional,sun2011nearly, tang2011hightemperature, neupert2011fractional,  yao2013realizing}. 
Since interactions on a lattice are intrinsically directional, FCIs can possess anisotropic PPs that are even larger than the isotropic ones.
The PP decomposition of an FCI requires a basis mapping between the single-body orbitals of the FCI and FQH systems \cite{qi2011generic,
jian2013crystal,lee2013pseudopotential,wu2012zoology,wu2013bloch,lee2014lattice}. We adopt the mapping introduced in Ref.\cite{claassen2015position}, which maps a Chern number $C$ Chern band to a \emph{single} LL with $C$-fold degeneracy \footnote{It thus avoids the complications of inter-LL mixings introduced by alternative mapping schemes (but see Ref. \onlinecite{wu2013bloch}).} without introducing basis anisotropy \cite{qi2011generic,jian2013crystal,lee2013pseudopotential,lee2014lattice}. For a clean realization of large PP FCIs, we choose a Chern insulator with almost uniform Berry curvature and Fubini-Study metric, so as to restrict rotational symmetry breaking to the interaction. Hence an interaction with predefined symmetry (i.e., $d$-wave) will only lead to anisotropic PPs of that symmetry. Using the 3-band, Chern number $3$ model \cite{SOM}, we plot in Fig.\ref{fig4_v3}(b) its PP coefficients computed via Eq. \ref{decompose2} for nearest neighbour interactions of the (unprojected) form $V_q=\cos q_x$ or $V_q=\cos q_x - \cos q_y$. In the former case, $V_{m,2}$ is of comparable magnitude with the isotropic PP $V_{m,0}$ (at least for $m=0$ and $1$), while in the latter (quadrupole) case, the isotropic PPs cancel and we are left only with anisotropic PPs.

 {\sl Metrics in generalized PPs.--} The existence of a geometric degree of freedom for the dynamics within a single LL is the starting point of recent theoretical approaches to the FQH effect \cite{HaldaneViscosity, HaldaneGeometry, QiuPhysRevB.85.115308, BoYangPhysRevB.85.165318,  XinWanPhysRevB.86.035122, Apalkov2014128, MaciejkoPhysRevB.88.125137}. The incompressibility of FQH phases in general does not require rotational symmetry, but model wavefunctions (e.g., Laughlin \cite{Laughlin-PhysRevLett.50.1395}, Moore-Read \cite{Moore1991362} states, etc.) \emph{are} rotationally invariant, and intrinsic metric in them appears as a ``hidden" variational parameter.
Quantum fluctuations of the metric determine the gap of the neutral excitations in the long-wavelength limit \cite{HaldaneGeometry,BoYang2012}, which is crucial for the incompressibility of the topological fluids. Geometric aspects of the compressible quantum Hall phases can be more easily probed experimentally, especially in higher LLs \cite{Lilly1999, Du1999}. 

While the importance of the metric in the many-body states have been recognized, the generalized PPs allow us to explicitly study the metric of the projected Hamiltonian of Eq.(\ref{h}). It is straightforward to extend the generalized PPs above to the case of a general metric $g$.  For convenience, we introduce a complex vector $\omega^a$ such that $\epsilon^{ab}\omega_a^*\omega_b=i$; then $g^{ab}=\omega^{a*}\omega^b+\omega^a\omega^{b*}$. Further, we define
\begin{eqnarray}\label{eq:metric}
\textbf{q}=\omega^aq_a, \;\;\; |q|^2 = g^{ab}q_aq_b.
\end{eqnarray}
With the above redefinitions, the expression for the PPs, Eqs. (\ref{g1}) and (\ref{g2}), as well as the orthogonality relations, continue to hold. This means that in general, we can expand any interaction as 
\begin{eqnarray}\label{expansion}
V_{\vec{q}}=\sum_{m,n,\sigma} c_{n,m}^{g\sigma} V_{n,m}^{g\sigma}(\vec{q}).
\end{eqnarray}
More generally, because of the $SO(2,1)$ invariance of the integration measure in the Hamiltonian (\ref{h}), we find that arbitrary two-body interactions are divided into equivalence classes where the members of the same family are interactions related to each other by a stretch and rotation of the guiding center metric. In other words, we can generate a family of interactions from the RHS of Eq. (\ref{expansion}) which have the same $\{ c_{n,m}^{g\sigma}\}$, but different PP metric $\eta$  in $V_{n,m}^{\eta\sigma}(\vec{q})$; all such interactions lead to the same spectrum upon substituting in Eq.(\ref{h}). For certain $V_{\vec{q}}$ that are anisotropic in the lab frame, one can tune the metric in Eq.(\ref{expansion}) to minimize the coefficients of the anisotropic PPs. More importantly, for gapped FQH fluids, we can truncate the expansion of the anisotropic interaction into the generalized PPs with appropriate metric. This would give new classes of anisotropic model Hamiltonians that could provide simpler description of topological phases and phase transitions where geometry plays an important role. 

The technical details on explicitly defining the variational metric in the PPs are outlined in \cite{SOM}. It can be shown that for gapped FQHE phases, the main effect of introducing particular anisotropy (i.e., by adding selected $V_{m,n}^{\pm}$) is to change the emergent metric of the gapped ground state, without significantly reducing the incompressibility gap. Similarly, tilting the magnetic field for a quantum Hall sample with a finite thickness can be shown to introduce rather small anisotropy, even with a very large in-plane magnetic field \cite{SOM}. 

{\sl Conclusions.--} We have formulated a notion of generalized PPs that completely describe any two-body effective interactions within a single LL. This allows for a systematic way to quantitatively describe anisotropic FQH systems, such as tilted magnetic field systems, nematic states and FCIs. The generalized PPs, possibly in combination with standard isotropic PPs, may give rise to new topological states with broken symmetry at finite filling factors. It would be interesting to study the dynamics of the few-particle ``molecular" bound states for purely anisotropic PPs in the limit of vanishing filling factor, as well as the many-body phases at large anisotropy which is ubiquitous in realistic FCI models.

{\sl Acknowledgements.} This work was supported in part by  Singapore A$^{\star}$STAR SERC ``Complex Systems" Research Programme grant 1224504056, and the Department of Energy, Office of Basic Energy Sciences
through Grant No. $\rm DE-SC0002140$. Zi-Xiang Hu was supported by NSFC No.1127403, 11674041 and FRF for the Central Universities No.CQDXWL-2014-Z006. 
Z.P. acknowledges support by EPSRC grant EP/P009409/1. Statement of compliance with EPSRC policy framework on research data: This publication is theoretical work that does not require supporting
research data.

\bibliography{pseudopotentials}
\clearpage 
\pagebreak

\onecolumngrid
\begin{center}
\textbf{\large Supplemental Online Material for ``Generalized Pseudopotentials for the Anisotropic Fractional Quantum Hall Effect" }\\[5pt]
\vspace{0.1cm}
\begin{quote}
{\small In this supplementary material we include some details on the metric as variational parameters for the generalized PPs and for the anisotropic quantum Hall systems, and on how to compute the the decomposition into generalized PPs for two physical systems mentioned in the main text: the quantum Hall system with a tilted magnetic field, and the FCI system where anisotropy is induced by the lattice structure. }\\[20pt]
\end{quote}
\end{center}
\setcounter{equation}{0}
\setcounter{figure}{0}
\setcounter{table}{0}
\setcounter{page}{1}
\setcounter{section}{0}
\makeatletter
\renewcommand{\theequation}{S\arabic{equation}}
\renewcommand{\thefigure}{S\arabic{figure}}
\renewcommand{\thesection}{S\Roman{section}}
\renewcommand{\thepage}{S\arabic{page}}
\vspace{1cm}
\twocolumngrid

\section{S1. The variational metric of the generalized pseudopotentials}

In the 2D manifold, the real space angular momentum is given by $\epsilon_{ac}g^{bc}r^ap_b$ ($r^a$ and $p_a$ are real space and momentum operators respectively), and the projected guiding center angular momentum is $\frac{1}{2}g_{ab}R^aR^b$. Thus the definition of the angular momentum requires a metric. For a lowest Landau level (LLL) Hamiltonian in which the guiding center angular momentum with the metric $g$ is a good quantum number, the system is still isotropic even when the metric is non-Euclidean and breaks rotational symmetry with $g\neq \mathbbm{1}$. But as long as the guiding center angular momentum is conserved, the metric $g$ is the metric which describes properties of eigenstates. For anisotropic Hamiltonians that contain more than one metric \cite{BoYangPhysRevB.85.165318}, the metric that characterizes the eigenstates, especially the gapped ground state, becomes an emergent quantity. This emergent metric in fact enters the definition of the generalized pseudopotentials (PPs), as we now show.

The intra-LL Hilbert space, as indexed by the eigenstates of the guiding center angular momentum, can be organized by the metric of the guiding center angular momentum operator. To see that, it is convenient to rewrite a unimodular metric $g$ as
\begin{eqnarray}
g_{ab}=\omega_a^*\omega_b+\omega_a\omega_b^*,
\end{eqnarray}
with $\omega_a$ as the complex vector satisfying $\epsilon^{ab}\omega_a^*\omega_b=i$. For the special case of $g^{ab}=\mathbbm{1}$, we have $\omega_x=1/\sqrt{2},\omega_y=i/\sqrt{2}$. A guiding center ladder operator (that raises or lowers the guiding center angular momentum) can thus be defined as $b=\omega_a^*R^a$, with $[b,b^\dagger]=1$. Correspondingly, in Eq.(3-5) of the main text we have
\begin{eqnarray}
 \textbf{q}=\omega^aq_a, \;\;\; |q|^2=g^{ab}q_aq_b.
\end{eqnarray}
With these redefinitions of $|q|^2$ and $\textbf{q}$, the formulas for generalized PPs remain the same as in the main text:
\begin{eqnarray}
V_{m,n}^{g+}\left(\vec q\right)&=& \lambda_n \mathcal{N}_{mn} \left(L_m^n\left(|q|^2\right)e^{-\frac{1}{2}|q|^2}\textbf{q}^n+c.c\right),\label{g1s}\\
V_{m,n}^{g-}\left(\vec q\right)&=& -i \mathcal{N}_{mn} \left(L_m^n\left(|q|^2\right)e^{-\frac{1}{2}|q|^2}\textbf{q}^n-c.c\right),\label{g2s}
\end{eqnarray} 
where we have inserted the superscript $g$ to emphasize the metric dependence $V_{m,n}^{g\sigma},\sigma=\pm$. The normalization factors $\mathcal{N}_{nm}$ and $\lambda_n$ are defined in the main text.

The orthogonality relations of the generalized PPs continue to hold for arbitrary metric $g$. Similarly, we can write the decomposition of any interaction $V_{\vec{q}}$ as
\begin{eqnarray}
V_{\vec{q}}&=&\sum_{m,n=0}^\infty \sum_{\sigma=\pm}   c^{g\sigma}_{m,n}V_{m,n}^{g\sigma}(\vec{q}),\label{decompose2s}\\
c^{g\sigma}_{m,n}&=&\int d^2q V_{\vec{q}} V_{m,n}^{g\sigma}\left(\vec q\right).\label{decompose2as}
\end{eqnarray}
Remember that $V_{\vec{q}}$ can contain multiple metrics and the system is anisotropic. According to Eq. (\ref{decompose2s}), such an interaction can be expanded in terms of the generalized PPs $V_{m,n}^{g\sigma}$ with an arbitrary unimodular metric $g$; physically, we can view $g$ as a variational parameter in Eq.(\ref{decompose2s}), and a different metric $g$ will lead to a different set of PP coefficients $c_{m,n}^{g\sigma}$. If we are able to find the metric that maximizes $c_1^g=c_{10}^{g+}$ with all other PPs being small, this metric is the emergent metric of the Laughlin state at $1/3$ filling factor; the emergent metric of other topological phases can be found analogously. The formulas Eq.(3), Eq.(4) and Eq.(5) in the main text are the special cases of Eq.(\ref{g1s}), Eq.(\ref{g2s}), and Eq.(\ref{decompose2s}) when we fixed $g^{ab}=\mathbbm{1}$. If $V_{\vec q}$ is rotationally invariant with metric $g$, then $c_{m,n}^{g\sigma}=0$ if $n\neq 0$, and the Haldane PPs form a complete basis.

For a specific $V_{\vec q}$ in Eq.(\ref{decompose2s}), if we define a coefficient vector $\vec c_g$ with $c_{m,n}^{g\sigma}$ as entries, it is a vector with the following length
\begin{eqnarray}\label{clength}
\norm{\vec c_g}^2=\int d^2qV_{\vec q}^2
\end{eqnarray}
that is independent of the metric $g$, assuming the integration does not diverge. The Hamiltonian of Eq.(1) of the main text, on the other hand, is invariant under an $SO(2,1)$ unitary transformation of the metric in the measure of the integration \cite{HaldaneViscosity}. If we plug $V_{\vec q}$ in Eq.(\ref{decompose2s}) into the projected Hamiltonian Eq.(1) with an arbitrary $g$, we can do a coordinate transformation as follows:
\begin{eqnarray}\label{coordinatetransform}
\mathcal{H} &=&\sum_{m,n=0}^\infty\sum_{\sigma=\pm}c_{m,n}^{g\sigma}\int d^2qV_{m,n}^{g\sigma}\left(\vec q\right) \bar \rho_{\vec q}\bar \rho_{-\vec q}\nonumber\\
&=&\sum_{m,n=0}^\infty\sum_{\sigma=\pm}c_{m,n}^{g\sigma}\int d^2q V_{m,n}^{\eta\sigma}\left(\vec q\right) \bar \rho_{\vec q}\bar \rho_{-\vec q}
\end{eqnarray}
where $\eta$ can be Euclidean. This implies that for each fixed $V_{\vec q}$ there is an equivalence class of effective interactions with $\vec c_g$ given by Eq.(\ref{decompose2s}) and length Eq.(\ref{clength}), parametrized by an arbitrary unimodular $g$. We can write the equivalence class to which $V_{\vec q}$ belongs as follows:
\begin{eqnarray}\label{eclass}
V_{\vec q,g}=\sum_{m,n=0}^\infty \left(c^{g+}_{m,n}V_{m,n}^{\eta+}\left(\vec q\right)+c^{g-}_{m,n}V_{m,n}^{\eta-}\left(\vec q\right)\right)
\end{eqnarray}
It consists of interactions $V_{\vec q,g}$ parametrized by $g$. The LHS of Eq.(\ref{eclass}) are basically interactions related to each other by a stretch and a rotation, leading to the same spectrum of the Hamiltonian.

While the equivalence class can be trivially detected when $V_{\vec q}$ contains a unique metric, it is less obvious when $V_{\vec q}$ is anisotropic -- two seemingly different interactions can be in the same equivalence class, essentially describing the same physical system. This is also important for the determination of the proper model Hamiltonians that are adiabatically connected to the realistic Hamiltonians with anisotropic $V_{\vec q}$: one can tune the metric on the RHS of Eq.(7) of the main text to maximise the coefficient of a specific set of PPs. 

To illustrate this, it is convenient to pick a parametrization of the vector $\omega^a$ as
\begin{eqnarray}
\omega^x = \frac{1}{\sqrt{2}} \left( \cosh \frac{\theta}{2} + e^{-i\phi} \sinh \frac{\theta}{2} \right), \\
\omega^y = \frac{-i}{\sqrt{2}} \left( \cosh \frac{\theta}{2} - e^{-i\phi} \sinh \frac{\theta}{2} \right). 
\end{eqnarray}
By previous definition, this gives a metric
\begin{eqnarray}\label{g}
g=\left(\begin{array}{ccc}
 \cosh \theta+\sinh \theta\cos \phi & \sin \phi\sinh \theta\\
 \sin \phi\sinh \theta & \cosh \theta-\sinh \theta\cos \phi\end{array}\right).
 \end{eqnarray}
With this, we can show by direct calculation the following interesting relationship:
 \begin{eqnarray}\label{relationship}
 \partial_\theta V_{1}^g=-\frac{\sqrt{3}}{2} \left( \cos\phi V_{1,2}^{g+} - \sin\phi V_{1,2}^{g-}\right).
 \end{eqnarray}
This relation is particularly useful for maximising $c_{1}^g$. It implies that the metric which maximises $c_{1}^g$ also leads to the vanishing of $c_{1,2}^{g+}$. 
Thus for an isotropic Hamiltonian of which the $\mu=1/3$ Laughlin state is the ground state, the leading correction of adding $V_{1,2}^{g\pm}$ is to alter the intrinsic metric of the Laughlin state \cite{HaldaneViscosity,QiuPhysRevB.85.115308}, as long as the spectrum remains gapped. We now illustrate this with the following model:  
 \begin{eqnarray}\label{example}
 V_{\vec q}=V_{1}^\eta+\lambda V_{1,2}^{\eta+}.
 \end{eqnarray}

\begin{figure}
\includegraphics[width=8cm]{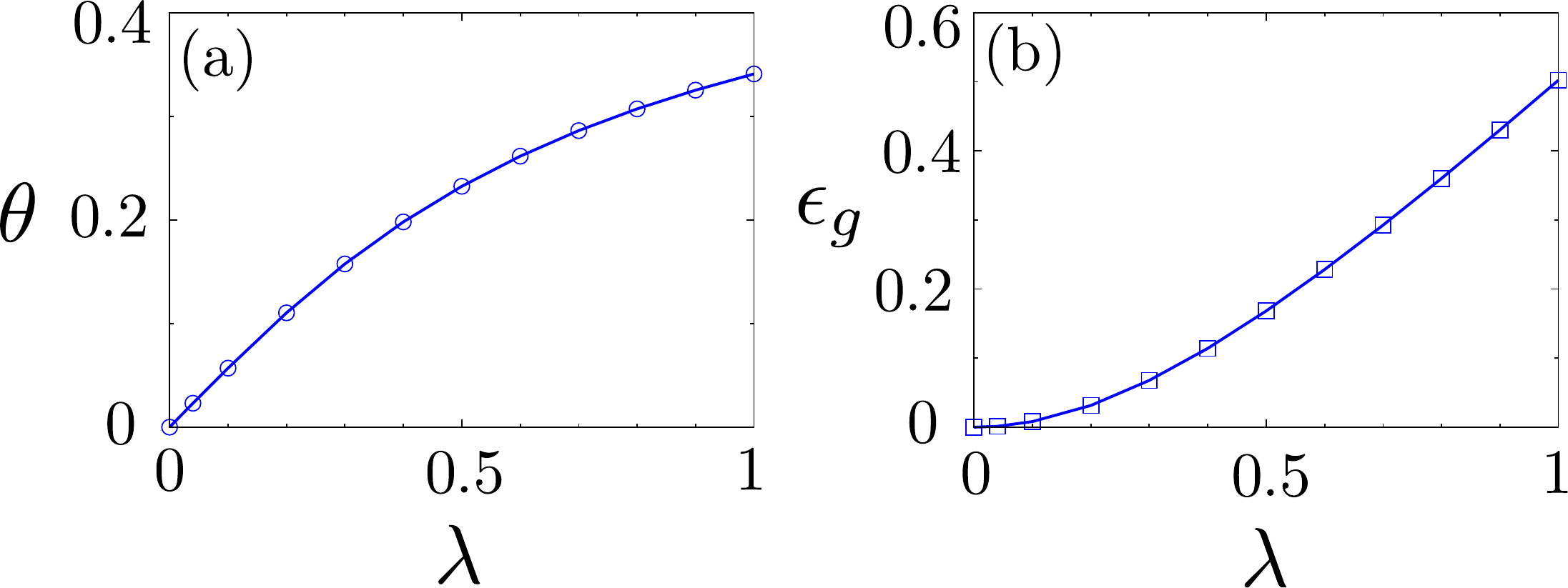}
\caption{(Color online) (a) The metric parameter $\theta$ which maximises $c_{10}^g$ as a function of $\lambda$ in Eq.(\ref{example}). (b) The dependence of the strength of perturbation on $\lambda$. Note that $\lambda$ is also the strength of perturbation of the equivalent interaction in Eq.(\ref{example}). By choosing the metric that maximises $c_{1}^g$, the strength of perturbation is substantially reduced, especially when $\lambda$ is small. } 
\label{correction}
\end{figure}

Without loss of generality we set the metric on the RHS to be $\eta=\mathbbm{1}$. The equivalent $V_{q,g}$ with the metric that maximises $c_{1}^g$ can be found variationally, which is pertinent at $\nu=1/3$ filling factor with various competing phases against the Laughlin state\cite{BoYangPhysRevB.85.165318, PapicTilt}. Formally we have
 \begin{eqnarray}\label{equivalent}
 V_{\vec q}\sim V_{\vec q,g}=c_{1}^gV_{1}^g+\sum'_{m,n}c^{g+}_{m,n}V_{m,n}^{g+}\left(\vec q\right)
 \end{eqnarray}
where the primed summation excludes $V_{1,2}^{g+}$ and $V_{1}^g$; the $V_{m,n}^{g-}$ are not present because of the symmetry of $V_{1,2}^{\eta+}$. Treating the second term of Eq.(\ref{example}) and Eq.(\ref{equivalent}) as the perturbation to the $1/3$ Laughlin state, the strength of perturbation of $V_{\vec q}$ is $\lambda$. Given that the coefficient vector is invariant under the unimodular transformation, the strength of perturbation of $V_{\vec q,g}$ is
\begin{eqnarray}\label{perturbation}
\epsilon_g=\sqrt{\sum'_{mn}\left(c_{m,n}^{g+}\right)^2}=\sqrt{1+\lambda^2-\left(c_{1,0}^g\right)^2}/c_{1,0}^g
\end{eqnarray}
The dependence of $\theta$ on $g$ and $\epsilon_g$ as a function of $\lambda$ are shown in Fig.(\ref{correction}). It is clear that the main effect of $V_{1,2}^{g\pm}$ is to modify the metric of the $1/3$ Laughlin liquid. In contrast, if one replaces $V_{1,2}^{g\pm}$ in Eq.(\ref{example}) with other anisotropic PPs, the metric that maximises the $V_{1}^g$ component remains as $\eta$ according to Eq.(\ref{relationship}), and the incompressibility gap will be reduced much more significantly.

\section{S2. Quantum Hall systems with tilted magnetic field and finite thickness}\label{tiltedfield}

\begin{figure}[htb]
\includegraphics[width=0.6\linewidth]{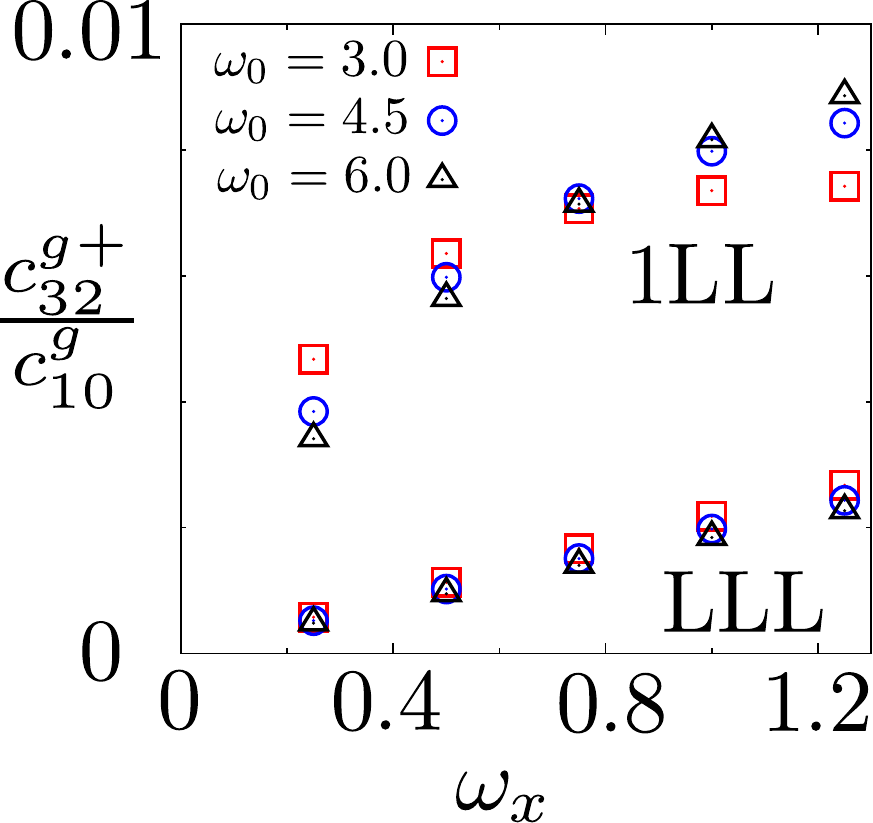}
\caption{(Color online) The magnitude of $V_{3,2}^{g+}$ induced by the in-plane field of cyclotron frequency $\omega_x$.  The coefficient of $V_{3,2}^{g+}$ is normalised by $c_{1}^{g}$, evaluated at the metric that maximises $c_{1}^g$. We show the results for $N=0$ (LLL) and $N=1$ (1LL) Landau level, and several thickness parameters $\omega_0$.}
\label{fig4_supp}
\end{figure}

We assume the quantum Hall system is confined to a two-dimensional sample with finite thickness. A strong magnetic field is applied perpendicular to the sample in the $z$-direction. To model the tilted field, we assume an additional component of the in-plane magnetic field is present along the $x$-direction. The single particle Hamiltonian is then given by:
\begin{eqnarray}\label{h1}
H&=&\frac{1}{2m}\left(\left(P_x+eA_x\right)^2+\left(P_y+eA_y\right)^2+\left(P_z+eA_z\right)^2\right)\nonumber\\
&&+\frac{1}{2}m\omega_0^2z^2
\end{eqnarray}
Here we model the finite thickness of the Hall sample with a harmonic well. We define the canonical momentum $\pi_i=P_i+eA_i$, with $i=1,2,3$ along directions $x,y,z$. We also define $\pi_4=m\omega_0z$. Thus Eq.(\ref{h1}) can be written as
\begin{eqnarray}\label{h2}
H=\frac{1}{2m}\left(\pi_1^2+\pi_2^2+\pi_3^2+\pi_4^2\right),
\end{eqnarray}
with the following commutation relationships
\begin{eqnarray}\label{commutation}
&&[\pi_1,\pi_2]=-i\ell_{B_z}^{-2},\nonumber\\
&&[\pi_2,\pi_3]=-i\ell_{B_x}^{-2},\nonumber\\
&&[\pi_3,\pi_1]=[\pi_1,\pi_4]=[\pi_2,\pi_4]=0,\nonumber\\
&&[\pi_3,\pi_4]=-i\ell_0^{-2},
\end{eqnarray}
where the three length scales are given by $\ell_{B_z}=1/\sqrt{eB_z},\ell_{B_x}=1/\sqrt{eB_x},\ell_0=1/\sqrt{m\omega_0}$. We can thus define two sets of coupled harmonic oscillators as follows
\begin{eqnarray}\label{ho1}
&&a=\frac{1}{\sqrt{2}}\ell_{B_z}\left(\pi_1-i\pi_2\right),\qquad b=\frac{1}{\sqrt{2}}\ell_0\left(\pi_3-i\pi_4\right),\nonumber\\
&&[a,a^\dagger]=[b,b^\dagger]=1,\nonumber\\
&&[a,b]=[a,b^\dagger]=-\frac{1}{2}\ell_{B_z}\ell_0\ell_{B_x}^{-2}.
\end{eqnarray}
The Hamiltonian can now be rewritten as 
\begin{eqnarray}\label{h3}
H=\frac{1}{2ml_{B_z}^2}\left(a^\dagger a+aa^\dagger\right)+\frac{1}{2ml_0^2}\left(b^\dagger b+bb^\dagger\right).
\end{eqnarray}
One can perform a proper Bogoliubov transformation to decouple the two sets of ladder operators and bring the Hamiltonian into the normal form.  It is important to choose a right set of parameters to deal with the transformation. Without loss of generality, we can set $\omega_z=1/\left(m\ell_{B_z}^2\right)=1$, so the parallel field and the confining potential strength are all measured with respect to the perpendicular field. We define the following parameters
\begin{eqnarray}\label{AB}
\epsilon_1&=&1+\omega_0^2+\omega_x^2,\qquad\epsilon_2=\omega_0^2,\\
\omega_1^2&=&\frac{1}{2}\left(\epsilon_1-\sqrt{\epsilon_1^2-\epsilon_2^2}\right),\\
\omega_2^2&=&\frac{1}{2}\left(\epsilon_1+\sqrt{\epsilon_1^2-\epsilon_2^2}\right),
\end{eqnarray}
where $\pm\omega_1$ and $\pm\omega_2$ are actually the eigenvalues, or the characteristic frequencies of the decoupled oscillators $X,X^\dagger$ and $Y,Y^\dagger$ in the Hamiltonian, and $\omega_x=1/\left(m\ell_{B_x}^2\right)$.The single particle Hilbert space is thus built from these two sets of decoupled ladder operators, which we label as $|m,n\rangle=\frac{1}{\sqrt{m!n!}}\left(X^\dagger\right)^m\left(Y^\dagger\right)^n|0\rangle$, $m,n$ are non-negative integers. The Landau level is now indexed by two integers 
$$|m,n\rangle=\frac{1}{\sqrt{m!n!}}\left(X^\dagger\right)^m\left(Y^\dagger\right)^n|0\rangle,$$ where $|0\rangle$ is the vacuum state. In Fig.(\ref{fig4_supp}), we are looking at the LLL state given by $|0,0\rangle$ and the 1LL state given by $|1,0\rangle$. The 1LL is the second lowest single particle energy state when $\omega_0>\omega_z$.

We now look at the density-density interaction Hamiltonian with a bare Coulomb interaction
\begin{eqnarray}\label{hint}
H_{\text{int}}=\int d^3qV_{\vec q}\rho_q\rho_{-q},
\end{eqnarray}
where $V_{\vec q}=1/q^2$ is the Fourier components of the Coulomb interaction, and $\rho_q=\sum_ie^{i\vec q\cdot r}$ is the density operator. Let us define the usual cyclotron and guiding center coordinates as follows:
\begin{eqnarray}\label{coordinates}
&&\tilde R^1=\ell_{B_z}^2\pi_2,\qquad\tilde R^2=-\ell_{B_z}^2\pi_1,\qquad\nonumber\\
&&\bar R^a=r^a-\tilde R^a, a=1,2,\nonumber\\
&&[\tilde R^a,\tilde R^b]=-[\bar R^a,\bar R^b]=-i\ell_{B_z}^2\epsilon^{ab},\nonumber\\
&&[\tilde R^a,\bar R^b]=0.
\end{eqnarray}
The part relevant to the Landau level form factor is thus given by $\langle m,n|e^{i\left(q_1\tilde R^1+q_2\tilde R^2+q_3r^3\right)}|m,n\rangle$. To get the form factor it is useful to define the following functions:
\begin{eqnarray}\label{function}
f_1\left(x,y\right)&=&\frac{\left(4+x\right)\left(4-y\right)\sqrt x}{16\left(x-y\right)},\nonumber\\
f_2\left(x,y\right)&=&\frac{\left(4-x\right)\left(4-y\right)\sqrt x}{32\left(x-y\right)},\nonumber\\
f_3\left(x,y\right)&=&\frac{\sqrt{x\left(4-x\right)\left(4-y\right)}}{4\sqrt 2\left(x-y\right)},\nonumber\\
f_4\left(x,y\right)&=&\frac{x-4}{\sqrt x\left(x-y\right)}.
\end{eqnarray}
The form factor is thus given by
\begin{eqnarray}
F_{mn}\left(\vec q, q_3\right)=\langle m,n|e^{i\left(q_1\tilde R^1+q_2\tilde R^2+q_3r^3\right)}|m,n\rangle.
\end{eqnarray}
Using the functions $f_i$, the form factor can be expressed as
\begin{widetext}
\begin{eqnarray}\label{form2}
&&F_{mn}\left(\vec q, q_3\right)=\nonumber\\
&&=\exp\{-\frac{1}{2}\left(\left(f_1\left(\omega_1^2,\omega_2^2\right)+f_1\left(\omega_2^2,\omega_1^2\right)\right)\textbf q\textbf q^* \right)\} \times \exp\{-\frac{1}{2}\left( \left(f_2\left(\omega_1^2,\omega_2^2\right)+f_2\left(\omega_2^2,\omega_1^2\right)\right)\left(\textbf q^2+\textbf q^{*2}\right) \right)\} \nonumber\\
&& \times \exp\{-\frac{1}{2}\left( \left(f_3\left(\omega_1^2,\omega_2^2\right)+f_3\left(\omega_2^2,\omega_1^2\right)\right)q_3\left(\textbf q-\textbf q^*\right) \right)\} \times \exp\{-\frac{1}{2}\left( \left(f_4\left(\omega_1^2,\omega_2^2\right)+f_4\left(\omega_2^2,\omega_1^2\right)\right)q_3^2\right)\}\nonumber\\
&& \quad\times\mathcal L_m\left(f_1\left(\omega_1^2,\omega_2^2\right)\textbf q\textbf q^*+f_2\left(\omega_1^2,\omega_2^2\right)\left(\textbf q^2+\textbf q^{*2}\right)+f_3\left(\omega_1^2,\omega_2^2\right)q_3\left(\textbf q-\textbf q^*\right)+f_4\left(\omega_1^2,\omega_2^2\right)q_3^2\right) \nonumber\\
&&\quad\times\mathcal L_n\left(f_1\left(\omega_2^2,\omega_1^2\right)\textbf q\textbf q^*+f_2\left(\omega_2^2,\omega_1^2\right)\left(\textbf q^2+\textbf q^{*2}\right)+f_3\left(\omega_2^2,\omega_1^2\right)q_3\left(\textbf q-\textbf q^*\right)+f_4\left(\omega_2^2,\omega_1^2\right)q_3^2\right),
\end{eqnarray}
\end{widetext}
where $\textbf q=\frac{1}{\sqrt{2}}\left(q_1-iq_2\right), \vec q=(q_1, q_2)$. The effective two-body interaction is thus given by $\tilde V_{\vec q, q_3}=\frac{1}{q^2}|F_{mn}(q)|^2$.
One still needs to integrate over $q_3$ to get the effective 2D interaction:
\begin{eqnarray}\label{finalvq}
V_{\vec q}=\int_{-\infty}^{\infty}dq_3\tilde V_{\vec q,q_3}F_{mn}^2.
\end{eqnarray}
The integration can be done numerically. We can now apply Eq.(\ref{decompose2as}) to generate the generalised PPs, which are plotted in Fig.(\ref{fig4_supp}). 

It should also be noted that in Fig.(\ref{fig4_supp}) we choose the metric as the one that maximises $c_{1}^g$, which is a sensible choice for understanding the ground states at $1/3$ filling factor and the phases competing with the Laughlin state \cite{BoYangPhysRevB.85.165318, PapicTilt}. At this metric we also have $c_{1,2}^{g\pm}=0$. We decompose the effective Coulomb interaction in the lowest LL and $N=1$ LL in terms of the dominant anisotropic PPs, i.e., we plot the coefficient of $V_{3,2}^{g+}$, normalised by $c_{1}^{g}$, as a function of $\omega_x$ for several values of $\omega_0$. Interestingly, the anisotropy induced by the tilt is actually rather small, even for very large tilting angles.

\section{S3. Details of the FCI model}
\label{FCI}

In this section, we show that large anisotropic PP components can exist in FCI systems and how we can construct them in a general way. Since interactions on a lattice are intrinsically directional, FCIs can possess anisotropic PPs that are even larger than the isotropic ones.


For an illustration of the PP decomposition of an anisotropic Fractional Chern Insulator, we use a next-nearest-neighbor (NNN)  $3$-band Chern number $C=3$ model engineered such that it is close to "ideal" - with an almost flat band and almost uniform Berry curvature and Fubini-Study metric. It has a large band flatness ratio (ratio of bandgap vs. bandwidth) of approximately $40$, mean-square fluctuation of the Berry curvature of $1.562\times 10^{-3}$ and mean-square fluctuation of the Fubini-Study metric of $1.09\times 10^{-2}$. Note that we require at least 3-bands to have an arbitrarily uniform Berry curvature.

The single-particle part of this FCI model consists of real-space hopping elements indexed by $H_{ij}(d_x,d_y)$, where $i,j=a,b,c$ indexes the orbitals (colors) and $d_x,d_y$ labels the unit cells at positions $d_x\hat x+d_y\hat y$. The hoppings in the $\hat x$ and diagonal directions are
\begin{equation}
H(1,0)=\left(
\begin{array}{ccc}
 6.65 & 4 i & -6 \\
 4 i & -4.87805 & 7.57576 i \\
 -6 & 7.57576 i & -1.78571
\end{array}
\right)\end{equation}
and
\begin{equation}H(1,1)=\left(
\begin{array}{ccc}
 -1.10833 & -3.125(1+i) & -3.5 i \\
 3.125(1-i) & -2.94118 & 1.66667(1+ i) \\
 3.5 i & -1.66667(1-i) & 4
\end{array}
\right)\end{equation}
Hoppings to the other NN and NNN unit cells are related by the following symmetries: Upon a spatial rotation of $\pi/2$ clockwise (i.e. $(1,1)\rightarrow (-1,1)$), $H_{aa},H_{bb},H_{cc}$ remained unchanged, $H_{ab},H_{bc},H_{ba},H_{cb}$ are multiplied by $-i$ and $H_{ac},H_{ca}$ are multiplied by $(-i)^2=-1$. 

\end{document}